\documentstyle[12pt,epsfig]{article}
\begin{document}
\begin{titlepage}
\begin{flushright}MPI-PhT/98-10
\end{flushright}
\begin{center}
\begin{huge}
NLO QCD Corrections 
and Triple Gauge Boson Vertices at the NLC \\
\end{huge}
\vspace{1cm}
K.J. Abraham \\
\vspace{.3cm}
~~~~Department of Physics,
University of Natal \\
~~Pietermaritzburg,
South Africa \\
\vspace{1cm}
Bodo Lampe~~~~ \\
\vspace{.3cm}
~~~Max Planck Institut f\"{u}r Physik,
F\"{o}hringer Ring 6 \\
D-80805 M\"{u}nchen,
Germany~~~\\
\vspace{.3cm}
\& \\
\vspace{.3cm}
Department of Physics,
University of Munich \\
Theresienstr. 37,
D-80833 M\"{u}nchen,
Germany \\
\vspace{1cm}
\begin{abstract} 
\noindent We study NLO QCD 
corrections as relevant to hadronic $W$ decay in $W$ pair 
production at a future 500 GeV $e^{+}e^{-}$ linac, with 
particular emphasis on 
the determination of triple gauge boson vertices. We find that 
hard gluon bremstrahlung may mimic signatures of
anomalous triple gauge boson vertices in certain distributions.
The size of these effects can strongly depend on the polarisation 
of the initial $e^{+}e^{-}$ beams.
\end{abstract}
\end{center}
\end{titlepage}

Although the Standard Model is in excellent agreement with existing
collider data, there are strong grounds 
to expect discrepancies to appear once 
future high energies accelerators 
are commissioned. In particular, Physics Beyond the Standard Model may
show up in the form of anomalous triple gauge boson vertices. Although
LEP data will constrain these trilinear couplings, high statistics 
at energies far above threshold are needed to pinpoint small deviations
from the Standard Model. Thus
detailed analyses have been performed on the sensitivity of
$W$ pair production and decay at future high energy $e^{+}e^{-}$ linacs 
\cite{schild2} \cite{snowmass} \cite{other}
to the presence of anomalous triple gauge boson vertices. 
These
studies, however, have not taken into account NLO QCD effects in hadronic
$W$ decay (see however \cite{pittau}). In this letter, we will argue that
these effects can generate deviations from tree level Standard Model 
predictions which are 
large enough to influence discovery bounds on anomalous form factors. 

The differential cross-section for $W$ pair production and decay 
(in the narrow width approximation which we use throughout
following \cite{schild1},\cite{schild2}) can be schematically 
written as 
\begin{equation}\label{eq:dsig}
d\sigma \sim \frac{\Gamma_{a}\Gamma_{b}}{\Gamma^{2}}
\sum_{\lambda \tau_{1} \tau_{2} \tau_{1}^{'} \tau_{2}^{'}} 
F^{\lambda}_{\tau_{1} \tau_{2}}(s,\cos{\vartheta})
{F^{\ast \lambda}_{\tau_{1}^{'} \tau_{2}^{'}}(s,\cos{\vartheta})}
D_{\tau_{1} \tau_{2}}
D_{\tau_{1}^{'} \tau_{2}^{'}}
\end{equation}

\noindent where $F$ is a generic helicity amplitude dependent on  
$\sqrt{s}$ and production angle $\vartheta$, and $D_{ab}$ denotes a 
generic element of the density 
matrix for $W$ decay. $\lambda$ denotes the electron helicity 
($\pm {{1}\over{2}}$) while
$\tau$ denotes $W^{\pm}$ helicities ($+$,$-$,$0$). $\Gamma$ is the 
total width of the $W$ while $\Gamma_{a}$ and $\Gamma_{b}$ denote partial
widths to the final states of interest. 
The precise forms
of $F$ and $D$ may be found in \cite{schild1}.
The constants we have not explicitly written 
in Eq.~\ref{eq:dsig} are purely kinematical overall factors common to all
final states and which are not relevant for the numerical results we will
present later on.

For the sake of convenience,
we reproduce the diagonal elements of $D$ from \cite{schild1}
\begin{eqnarray}\label{eq:LO} 
D_{++} =& \frac{1}{2}(1 + {\cos^{2}{\theta}})& -  \{\cos{\theta}\} \\
D_{--} =& \frac{1}{2}(1 + {\cos^{2}{\theta}})&  + \{\cos{\theta}\} \\
D_{00} =& {\sin^{2}{\theta}} & 
\end{eqnarray}
where $\theta$ is the polar angle of the outgoing fermion
in the rest frame
of the decaying $W$ and all fermions are assumed massless. 
We will assume that the rest frame of each $W$ can 
be reconstructed, thereby excluding purely leptonic final states. For
hadronic $W$ decays, where the jet charges cannot be reconstructed, 
symmetrisation between quarks and anti-quarks requires that the
terms within \{\} must be dropped. The off diagonal elements of $D$ 
depend, in addition to $\theta$, on the azimuthal angle $\phi$, but
make no contribution to the total cross-section. They are however
relevant for azimuthal correllations, which we will not discuss.

Once NLO QCD effects are included in $W$ decays, the formulae above must 
be modified. In addition to $\theta$ and $\phi$, the matrix elements for
gluon bremstrahlung contributions depend on additional phase space 
variables, and are singular in the collinear limit, quite apart from
divergences due to virtual corrections. As these singularities cancel 
only when IR safe quantities are calculated, IR safe generalisations of 
$\theta$ and $\phi$ are required. One way of proceeding follows from the 
observation that $\theta$ in Eq.~\ref{eq:LO}, defined in terms of the 
quark direction, is at LO also the polar angle of the thrust direction.
As thrust is IR safe \cite{Farhi}, NLO $D$ functions defined in terms of 
the thrust
orientation, which for our purposes is the direction of the most 
energetic outgoing parton in the $W$ rest frame, are gaurenteed 
to be singularity free.

Restricting ourselves once again to diagonal matrix elements we have at
NLO \cite{NPB} 
\begin{equation}\label{eq:HO}
D_{AA} = (1 + \frac{\alpha_{s}}{\pi})(1 -3LC_{F}\frac{\alpha_{s}}{2\pi}) 
\times\{D^{0}_{AA} + 2LC_{F}\frac{\alpha_{s}}{2\pi}\}
\end{equation}
neglecting terms of ${\cal O}(\alpha_{s}^{2})$. 
$D^{0}$ denotes the symmetrised leading order term and $L$ is 
a numerical constant 
of value $.4875$ which is a relic of numerical integration over Dalitz 
variables. All angles in Eq.~\ref{eq:HO} now refer not to a given 
outgoing parton but to the thrust axis. As before, all outgoing partons
are assumed massless. It should be noted that there are additional 
terms linear in $\cos{\theta}$ in Eq.~\ref{eq:HO} which 
vanish once we assume that jet charges are not determined, and which we
have therefore dropped.

Retaining terms to ${\cal O}(\alpha_{s})$ only 
\begin{displaymath}
\sum_{A} D_{AA} = (1 + \frac{\alpha_{s}}{\pi})\sum_{A} D^{0}_{AA}
\end{displaymath}
The term in brackets is the well known ${\cal O}(\alpha_{s})$ 
QCD K factor for hadronic W decay, as expected. We thus see, that 
for observables for which polarisation is not relevant, the NLO
corrections may be obtained by rescaling LO results by a constant
factor which we will derive shortly. In the Standard Model where 
gauge cancellations 
ensure the suppression of longitudinal gauge boson production 
at high energies, polarisation may be relevant for some observables,
for example asymmetries, where different polarisation states are
in general weighted differently. Hence we will not 
only approximate NLO effects by a constant K factor, but will also make 
use of Eq.~\ref{eq:HO} convoluted with Eq.~\ref{eq:dsig}. 

Before proceeding further along these lines, it is useful to study the
possible significance of NLO QCD effects in the analysis of anomalous
gauge boson vertices. To do so, it is instructive to 
consider the term within 
\{\} in Eq.~\ref{eq:HO} for $A = +$. This term can be rewritten as
\begin{displaymath}
\frac{(1 + \cos^{2}{\theta})}{2}(1 + 2LC_{F}\frac{\alpha_{s}}{2\pi}) + 
\frac{\sin^{2}{\theta}}{2}2LC_{F}\frac{\alpha_{s}}{2\pi}
\end{displaymath}
For transversely polarised gauge bosons, the LO distribution gets rescaled 
and in addition a longitudinally polarised component seems to appear. 
However, the appearance of additional longitudinal modes is one of the
hallmarks of non-standard triple gauge boson vertices ! Thus we see that
NLO QCD has the potential to mimic signatures of anomalous triple gauge
boson vertices.
The size of this effect is proportional to $L$, indicating
that hard gluon bremstrahlung is responsible. It is significant that
in Eq.~\ref{eq:HO} $\alpha_{s}$ is evaluated at $M_{W}$ independent 
of $\sqrt{s}$; NLO QCD corrections thus do not diminish in size with
increasing energy, in contrast to other radiative corrections such
as initial state bremstrahlung and finite width contributions.

For longitudinal modes ($A = 0$) it is easy to see from Eq.~\ref{eq:LO} 
and Eq.~\ref{eq:HO} that the QCD K factor at $\theta = 0$ is infinite.
Thus large QCD corrections may be expected in distributions where 
longitudinal modes are important. The infinite K factor is due to the
vanishing of the LO cross-section at $\theta = 0$, which may be understood
in terms of angular momentum conservation, and has been observed in other
processes involving hadronic $W$ decay \cite{schmidt}.

To derive expected $K$ factors, it is important to note that 
NLO QCD effects also
appear in the redefinition of the width and branching fractions of the $W$, 
which appear in Eq.~\ref{eq:dsig}. As we will 
focus on
final states containing leptons and hadrons, what appears at Born level
is (in obvious notation), $
\frac{\Gamma_{L}\Gamma_{H}}{\Gamma^{2}} $ where all widths
are calculated from tree level expressions. Keeping terms up to and 
including ${\cal O}(\alpha_{s})$ only \footnote{This factor is set to
1 in \cite{NPB} due to a different choice of overall normalisation.}
\begin{displaymath}
\Gamma^{2} \rightarrow \Gamma^{2}(1 + 2\frac{2}{3}\frac{\alpha_{s}}{\pi})
\end{displaymath}
The NLO cross-section with no phase space cuts can be obtained from Born level
by making a further change {\em i.e.}
\begin{displaymath}
\Gamma_{H} \rightarrow (1 + \frac{\alpha_{s}}{\pi})\Gamma_{H}
\end{displaymath}
Thus the change to the cross-section can be accounted for by rescaling
by a factor $\overline{K}$ given by \begin{equation}\label{eq:Kfac}
\overline{K} = (1 - \frac{1}{3}\frac{\alpha_{s}}{\pi})
\end{equation}
neglecting terms with higher powers of $\alpha_{s}$. This is the constant
factor mentioned earlier.

As we have argued earlier, the sum total of NLO QCD corrections may not 
always be obtained by rescaling by the constant factor in Eq.~\ref{eq:Kfac}.
Therefore we will define a $C$, a modified $K$ factor as follows;
\begin{equation}\label{eq:Cfac}
C = \frac{(NLO - \overline{K}LO)}{\overline{K}LO}
\end{equation}
If $C$ vanishes or is very small for a certain observable in a given 
region of phase space, then a rescaling is sufficient to describe 
NLO effects. If this is not the case, then NLO effects are significant.
Similarly, we can define $C_{A}$, to describe the corrections due to
anomalous couplings as follows, \begin{equation}\label{eq:Afac}
C_{A} = \frac{(A - LO)}{LO}
\end{equation} where $A$ is the contribution for a certain 
choice of anomalous triple gauge boson vertices.

We will now present numerical results for various observables at LO 
and NLO to illustrate the size of corrections to be expected. We fix 
throughout, $\sqrt{s} = 500\;GeV$, $M_{Z} = 91.187\;GeV$,
$M_{W} = 80.33\;GeV$, $\alpha = {1\over 128}$, $\sin{\theta_{W}}^{2}
= .23$, and $\alpha_{s} = .12$. For the sake of definiteness, we 
assume that the $W^{-}$ decays hadronically, and the $W^{+}$ 
leptonically. The incoming $e^{+}e^{-}$ beams are unpolarised unless 
otherwise stated, and the anomalous contributions are defined via 
Eq.~\ref{eq:Afac}. Angles referring to outgoing fermions and jets are
defined in the rest frame of the parent $W^{\pm}$.

To begin with, we demonstrate the utility of absorbing $\overline{K}$
into the LO cross-section by studying the $C$ 
dependence of the differential cross-section with respect to $\theta_{-}$,
the polar angle of the thrust axis. The two curves in Fig. 1 correspond
to $C$ with $\overline{K}$ defined by Eq.~\ref{eq:Kfac} and 
$\overline{K} = 1$. 
We see immediately 
that in this instance a significant part of the NLO corrections can be 
absorbed into the
redefinition of widths, thereby reducing the magnitude of such effects.

However, this is not always the case. For example,
for observables for which the LO contributions vanish in certain
regions of phase space, NLO effects cannot be simply accounted for by 
a redefinition of widths.
One such observable discussed in \cite{NPB} is given
by \begin{equation}\label{eq:zz}
\int_{-1}^{1} d\cos {\theta_{+}} 
\frac{\cos {\theta_{+}}}{|\cos {\theta}_{+}|} 
\frac{{d \sigma(e^{+}e^{-} \rightarrow \ell^{+} \nu  j_{-}X)}}
{d\cos{\vartheta} d\cos {\theta_{-}} d\cos {\theta_{+}} }
\end{equation}
which corresponds to the double differential distribution 
with respect to $\cos{\vartheta}$ and $\cos{\theta_{-}}$, with the 
azimuthal angles integrated over and the polar angle of the charged
anti-lepton integrated over anti-symmetrically.

This observable may seem rather contrived, however, being asymmetric by
construction it is sensitive to the C and P Violating form factor denoted
by $z_{z}$ in \cite{schild1}. As can be seen from Table 1, 
NLO QCD effects and a non zero value of $z_{z}$ both generate appreciable
corrections to the Standard Model predictions for the distribution 
described in Eq.~\ref{eq:zz}, particularly for $\cos{\vartheta} \sim 0$.
The need for taking into account NLO QCD effects in the
analysis of anomalous triple gauge boson vertices is apparent.

As has been pointed out in \cite{schild1} and \cite{schild2}, triple
differential distributions are particularly sensitive to anomalous gauge 
boson couplings. Hence, 
as a further illustration of the relevance of Eqs.~\ref{eq:HO} we now
sample a triple differential distribution with 
all azimuthal angles integrated over and the polar angle of the 
thrust axis is fixed at 0.1. 
In addition, the incoming beams are polarised ($e^{-}_{R}$ and
$e^{+}_{L}$).
The non-zero anomalous couplings are 
choosen to be (in the notation of \cite{schild1}) \begin{displaymath}
x_{\gamma} = .005 \;\;
\delta_{z} =  \frac{x_{\gamma}}{\sin{\theta_{W}}\cos{\theta_{W}}}\;\;
x_{z} =  -x_{\gamma}\frac{\sin{\theta_{W}}}{\cos{\theta_{W}}}
\end{displaymath}
This choice of parameters is motivated by the scenario in \cite{malaampi}
and the values above are slightly above the threshold for discovery at 
$\sqrt{s} = 500\,GeV$ according to LO analyses in \cite{schild2}.   

>From the results in Table 2 it is clear that although the anomalous
couplings produce sizable deviations from the tree level predictions 
of the Standard Model, NLO QCD effects are definitely not negligible in
comparision and need to be taken into account to establish discovery
limits. 

It is worth noting that for opposite incoming beam helicities, both 
the anomalous corrections and NLO QCD corrections in the same
region of phase space are much smaller; the NLO QCD corrections are 
never more than a few percent. This can be understood from the fact
that for incoming $e^{-}_{R}$, the outgoing $W^{-}$ is largely 
longitudinally polarised, while for incoming $e^{-}_{L}$ the outgoing
$W^{-}$ is largely transverse. 
This strong dependence of the size of NLO QCD effects on the 
incoming beam polarisation has not been
pointed out before, and is particularly significant, as several authors
have suggested beam polarisation as a diagnostic tool to unravel the
structure of anomalous gauge boson interactions \cite{snowmass}
\cite{dawval}. It is also noteworthy that Tables 1 \& 2 
are so different from each other, indicating that NLO QCD corrections
to different observables may not be simply obtained from some universal
prescription, but must be calculated from scratch.

Common to the results presented in Tables 1 and 2, apart from the strong
dependence of K factors with phase space, is the fact that different 
polarisation states make different contributions to the distributions under
consideration, either due to a choice of initial polarisation or due to an
observable being asymmetric by construction. As we have argued earlier
it is precisely in such cases that NLO QCD corrections could be non-trivial
and this is indeed consistent with our results, and with Figs. 1-3 of
\cite{NPB} where a sizable variation in K factors is also observed.
This suggests a useful rule
of thumb; higher order QCD corrections should not be approximated by constant
K factors where $W$ polarisation is observed and/or where 
different polarisation states make different contributions 
to the observables under consideration. Finally, it is worth repeating that 
the magnitude of the relevant K factors is controlled by 
$\alpha_{s}(M_{W})$, and is 
thus independent of $\sqrt{s}$ insofar as the narrow width approximation
is valid. It is not surprising therefore, 
that the broad features of Tables 1 and 2 persist at higher energies as well. 

To summarise, we have demonstrated the importance of NLO QCD corrections 
in the analysis of triple gauge boson vertices at future $e^{+}e^{-}$ 
linacs. The magnitude of these corrections varies strongly with beam 
polarisation and seem to be particularly large for
asymmetries, and certainly will affect the exclusion bounds for anomalous
triple gauge boson vertices. A precise quantitative estimate will be 
possible only with a detailed analysis including detector acceptances
which is beyond the scope of this letter, but is definitely worth 
undertaking.

\noindent {\em Acknowledgements} KJA wishes to thank Jose Wudka for 
valuable clarifications and encouragement. BL acknowledges useful
discussions with J.G. K\"{o}rner.

\newpage
\noindent{\bf Captions} \\
\vspace{1cm}
\\
\noindent {\em Figure 1} \\
\\
\noindent The variable $C$ with $\overline{K}$ defined in 
Eq.~\ref{eq:Kfac} (dashed line) and $\overline{K} = 1$ (solid line)
are plotted as a function of $\theta_{-}$, the polar angle of the thrust 
axis, with all other angles integrated over. 
$\theta_{-}$ is retricted by 
$0 < \theta_{-} < \frac{\pi}{2}$ as jet charge is assumed not to be
identified. 
\\
\vspace{1cm}
\\
\noindent {\em Table 1} \\
\noindent The values of $C$ and $C_{A}$ are plotted to three significant 
figures as a 
function of $\cos{\vartheta}$ and $\theta_{-}$ for the variable defined in 
Eq.~\ref{eq:zz};the upper figure in each entry is $C$, defined 
the text, while the lower figure is $C_{A}$ evaluated for $z_{z} = .001$, 
with all other anomalous parameters set to 0. $\cos{\vartheta}$ runs along 
the vertical axis and $\theta_{-}$ (in units of $\pi$) along the 
horizontal axis. 
\\
\vspace{1cm}
\\
\noindent {\em Table 2} \\
\noindent The values of $C$ and $C_{A}$ are plotted as a function of 
$\cos{\vartheta}$ and $\theta_{+}$ for the double differential 
cross-section evaluated at $\theta_{-} = .1$ with polarised beams
($e^{-}_{R}$ and $e^{+}_{L}$). 
The upper figure in each entry is $C$, while the lower figure is 
$C_{A}$ evaluated for anomalous parameters described in the text.
$\cos{\vartheta}$ runs along 
the vertical axis and $\theta_{+}$ (in units of $\pi$) along the 
horizontal axis. All azimuthal angles have been integrated over.

\newpage

\begin{figure}
\begin{center}
\epsfig{file=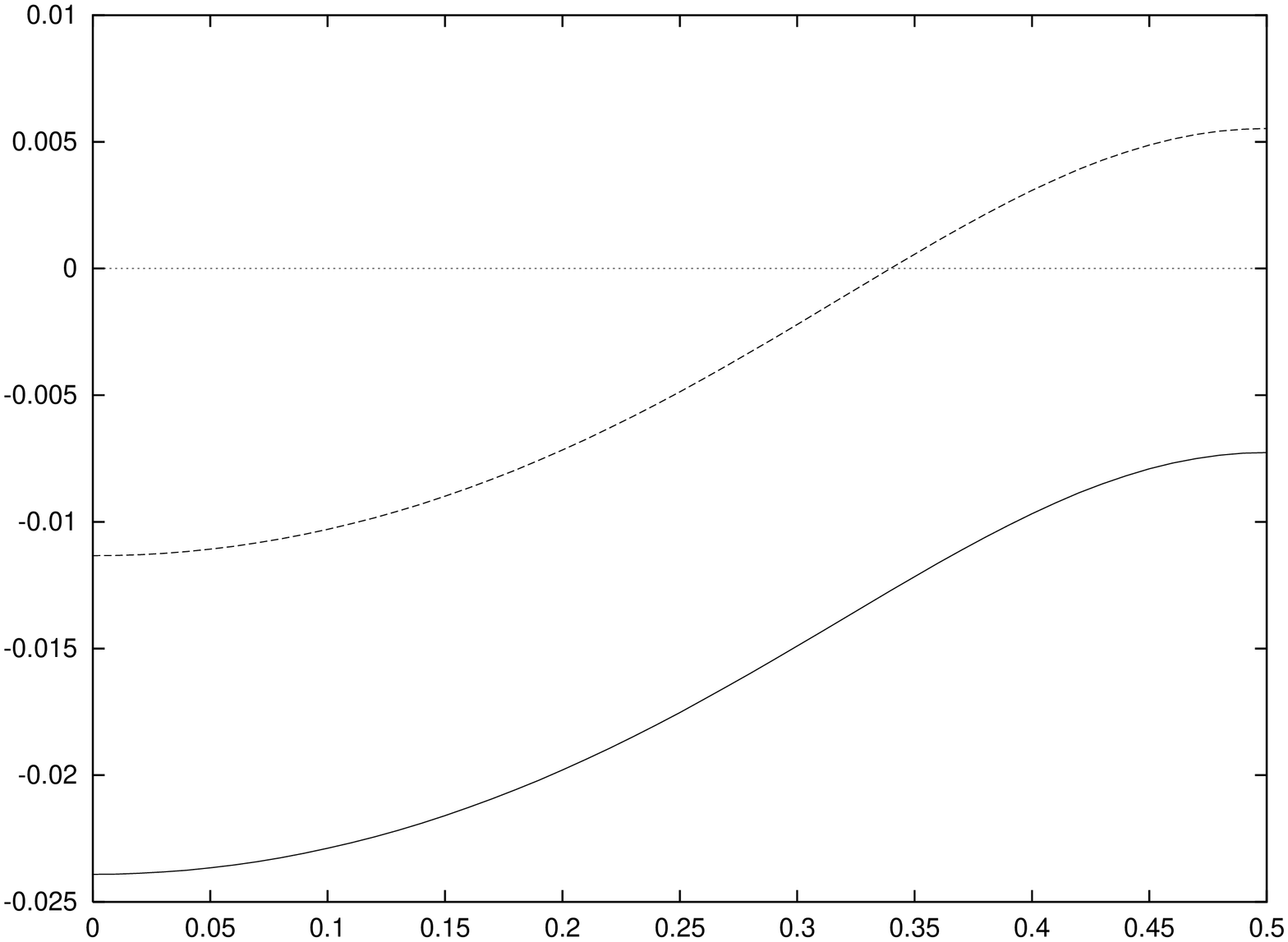,height=15cm,angle=270}
\bigskip
\bigskip
\caption{}
\end{center}
\end{figure}

\newpage

\noindent Table 1
\newline
\begin{tabular}{l|l|l|l|l|l|l|l|l|l|l|l|l}
\vspace{.2cm}
    & 0   &.05  &.10  &.15  &.20  &.25  &.30  &.35  &.40  &.45  &.50 \\ \hline
-.8 &  -.012&-.012&-.011&-.010&-.008&-.005&-.002& .001& .004& .006& .007\\
 &   .000& .001& .003& .007& .012& .019& .027& .036& .044& .050& .052\\ \hline
-.7 &  -.014&-.014&-.013&-.011&-.009&-.006&-.002& .002& .006& .008& .010\\
 &   .000& .000& .002& .004& .007& .011& .016& .022& .027& .032& .033\\ \hline
-.6 &  -.015&-.015&-.014&-.012&-.010&-.006&-.002& .002& .007& .010& .011\\
 &   .000& .000& .001& .003& .005& .008& .011& .015& .019& .022& .023\\ \hline
-.5 &  -.016&-.015&-.014&-.013&-.010&-.007&-.002& .003& .008& .011& .013\\
 &   .000& .000& .001& .002& .003& .006& .008& .012& .015& .017& .018\\ \hline
-.4 &  -.016&-.016&-.015&-.013&-.011&-.007&-.002& .003& .009& .013& .015\\
 &   .000& .000& .001& .002& .003& .005& .007& .010& .013& .016& .017\\ \hline
-.3 &  -.017&-.017&-.016&-.014&-.011&-.007&-.002& .004& .010& .015& .017\\
 &   .000& .000& .001& .002& .003& .005& .008& .011& .015& .018& .019\\ \hline
-.2 &  -.019&-.018&-.017&-.015&-.012&-.008&-.003& .005& .013& .020& .023\\
 &   .000& .000& .001& .002& .004& .007& .011& .017& .022& .028& .030\\ \hline
-.1 &  -.023&-.022&-.021&-.019&-.016&-.011&-.003& .010& .028& .050& .061\\
 &   .000& .000& .002& .005& .009& .016& .028& .045& .071& .101& .117\\ \hline
0. & $\infty$ & .973& .220& .081& .033& .010&-.001&-.008&-.012&-.014&-.014\\
 &     &-.061&-.061&-.061&-.061&-.061&-.061&-.061&-.061&-.061&-.061\\ \hline
.1 &  -.007&-.007&-.006&-.006&-.005&-.003&-.002&-.001& .000& .001& .001\\
 &   .000&-.001&-.002&-.004&-.008&-.011&-.015&-.019&-.022&-.025&-.025\\ \hline
.2 &  -.011&-.011&-.010&-.009&-.007&-.005&-.002& .000& .003& .004& .005\\
 &   .000& .000&-.001&-.002&-.004&-.006&-.009&-.012&-.014&-.016&-.016\\ \hline
.3 &  -.012&-.012&-.011&-.010&-.008&-.005&-.002& .001& .004& .006& .007\\
 &   .000& .000&-.001&-.002&-.003&-.004&-.006&-.008&-.010&-.012&-.012\\ \hline
.4 &  -.013&-.013&-.012&-.010&-.008&-.006&-.002& .001& .005& .007& .008\\
 &   .000& .000&-.001&-.001&-.002&-.003&-.005&-.006&-.008&-.009&-.009\\ \hline
.5 &  -.014&-.013&-.012&-.011&-.009&-.006&-.002& .002& .005& .008& .009\\
 &   .000& .000& .000&-.001&-.002&-.003&-.004&-.005&-.006&-.007&-.007\\ \hline
.6 &  -.014&-.014&-.013&-.011&-.009&-.006&-.002& .002& .005& .008& .009\\
 &   .000& .000& .000&-.001&-.001&-.002&-.003&-.004&-.005&-.005&-.005\\ \hline
.7 &  -.014&-.014&-.013&-.011&-.009&-.006&-.002& .002& .006& .009& .010\\
 &   .000& .000& .000& .000&-.001&-.001&-.002&-.003&-.003&-.004&-.004\\ \hline
.8 &  -.014&-.014&-.013&-.011&-.009&-.006&-.002& .002& .006& .009& .010\\
 &   .000& .000& .000& .000& .000&-.001&-.001&-.002&-.002&-.002&-.002\\ \hline
\end{tabular}
\newpage
\noindent Table 2
\newline
\begin{tabular}{l|l|l|l|l|l|l|l|l|l|l|l|l}
\vspace{.2cm}
    & 0   &.05  &.10  &.15  &.20  &.25  &.30  &.35  &.40  &.45  &.50 \\ \hline
  -.8 &1.551& .874& .393& .218& .144& .106& .085& .072& .064& .058& .054\\
   & .290& .252& .225& .216& .212& .210& .209& .208& .208& .207& .207\\ \hline
  -.7 &1.275& .783& .389& .234& .167& .133& .113& .101& .093& .087& .084\\
   & .314& .269& .233& .219& .212& .209& .208& .207& .206& .206& .206\\ \hline
  -.6 &1.062& .701& .382& .249& .189& .158& .140& .130& .122& .117& .114\\
   & .333& .284& .240& .221& .213& .209& .207& .205& .205& .204& .204\\ \hline
  -.5 & .892& .627& .373& .261& .209& .182& .167& .157& .151& .146& .143\\
   & .348& .296& .246& .224& .214& .209& .206& .204& .203& .203& .203\\ \hline
 -.4& .753& .559& .360& .269& .226& .204& .191& .182& .177& .173& .171\\
   & .360& .307& .252& .226& .215& .209& .205& .203& .202& .201& .201\\ \hline
-.3 & .638& .496& .345& .273& .239& .221& .210& .204& .199& .197& .194\\
   & .371& .315& .256& .229& .215& .208& .204& .202& .201& .200& .200\\ \hline
-.2   & .541& .439& .327& .273& .247& .233& .225& .220& .217& .214& .213\\
   & .379& .322& .260& .230& .216& .208& .204& .201& .200& .199& .199\\ \hline
 -.1  & .458& .385& .306& .267& .248& .238& .233& .229& .227& .226& .225\\
   & .387& .328& .263& .231& .216& .208& .204& .201& .199& .199& .198\\ \hline
 0.  & .386& .336& .282& .256& .243& .237& .233& .231& .230& .229& .229\\
   & .393& .331& .264& .232& .216& .208& .203& .201& .199& .198& .198\\ \hline
.1   & .323& .290& .256& .240& .232& .228& .226& .225& .224& .224& .225\\
   & .399& .333& .264& .231& .216& .208& .204& .201& .199& .199& .198\\ \hline
.2   & .268& .248& .227& .218& .214& .212& .212& .211& .211& .212& .213\\
   & .403& .333& .263& .231& .216& .208& .204& .201& .200& .199& .199\\ \hline
 .3  & .219& .208& .198& .193& .192& .191& .191& .191& .192& .193& .194\\
   & .408& .331& .260& .229& .215& .208& .204& .202& .201& .200& .200\\ \hline
.4  & .175& .170& .167& .165& .165& .165& .166& .166& .167& .169& .171\\
   & .412& .326& .256& .227& .215& .209& .205& .203& .202& .201& .201\\ \hline
.5   & .135& .135& .135& .135& .136& .136& .137& .138& .139& .141& .143\\
   & .415& .319& .250& .225& .214& .209& .206& .204& .203& .203& .203\\ \hline
 .6  & .100& .102& .103& .104& .105& .106& .107& .108& .109& .111& .114\\
   & .418& .309& .244& .222& .213& .209& .207& .205& .205& .204& .204\\ \hline
 .7  & .067& .071& .072& .073& .074& .075& .076& .077& .079& .081& .084\\
   & .421& .294& .236& .219& .213& .209& .208& .207& .206& .206& .206\\ \hline
.8   & .038& .041& .042& .043& .043& .044& .045& .047& .048& .051& .054\\
   & .424& .274& .228& .216& .212& .210& .209& .208& .208& .207& .207\\ \hline
\end{tabular}
\newpage
\noindent Table 2 {\em cont.}
\newline
\begin{tabular}{l|l|l|l|l|l|l|l|l|l|l|l|l}
\vspace{.2cm}
    & .5   &.55  &.6  &.65  &.7  &.75  &.80  &.85  &.90  &.95  &1. \\ \hline
  -.8 & .054& .051& .048& .047& .045& .044& .043& .043& .042& .041& .038\\
   & .207& .207& .208& .208& .209& .210& .212& .216& .228& .274& .424\\ \hline
  -.7 & .084& .081& .079& .077& .076& .075& .074& .073& .072& .071& .067\\
   & .206& .206& .206& .207& .208& .209& .213& .219& .236& .294& .421\\ \hline
 -.6  & .114& .111& .109& .108& .107& .106& .105& .104& .103& .102& .100\\
   & .204& .204& .205& .205& .207& .209& .213& .222& .244& .309& .418\\ \hline
 -.5  & .143& .141& .139& .138& .137& .136& .136& .135& .135& .135& .135\\
   & .203& .203& .203& .204& .206& .209& .214& .225& .250& .319& .415\\ \hline
 -.4  & .171& .169& .167& .166& .166& .165& .165& .165& .167& .170& .175\\
   & .201& .201& .202& .203& .205& .209& .215& .227& .256& .326& .412\\ \hline
 -.3  & .194& .193& .192& .191& .191& .191& .192& .193& .198& .208& .219\\
   & .200& .200& .201& .202& .204& .208& .215& .229& .260& .331& .408\\ \hline
 -.2  & .213& .212& .211& .211& .212& .212& .214& .218& .227& .248& .268\\
   & .199& .199& .200& .201& .204& .208& .216& .231& .263& .333& .403\\ \hline
 -.1  & .225& .224& .224& .225& .226& .228& .232& .240& .256& .290& .323\\
   & .198& .199& .199& .201& .204& .208& .216& .231& .264& .333& .399\\ \hline
 0. & .229& .229& .230& .231& .233& .237& .243& .256& .282& .336& .386\\
   & .198& .198& .199& .201& .203& .208& .216& .232& .264& .331& .393\\ \hline
  .1 & .225& .226& .227& .229& .233& .238& .248& .267& .306& .385& .458\\
   & .198& .199& .199& .201& .204& .208& .216& .231& .263& .328& .387\\ \hline
 .2  & .213& .214& .217& .220& .225& .233& .247& .273& .327& .439& .541\\
   & .199& .199& .200& .201& .204& .208& .216& .230& .260& .322& .379\\ \hline
 .3  & .194& .197& .199& .204& .210& .221& .239& .273& .345& .496& .638\\
   & .200& .200& .201& .202& .204& .208& .215& .229& .256& .315& .371\\ \hline
 .4  & .171& .173& .177& .182& .191& .204& .226& .269& .360& .559& .753\\
   & .201& .201& .202& .203& .205& .209& .215& .226& .252& .307& .360\\ \hline
 .5  & .143& .146& .151& .157& .167& .182& .209& .261& .373& .627& .892\\
   & .203& .203& .203& .204& .206& .209& .214& .224& .246& .296& .348\\ \hline
 .6  & .114& .117& .122& .130& .140& .158& .189& .249& .382& .701&1.062\\
   & .204& .204& .205& .205& .207& .209& .213& .221& .240& .284& .333\\ \hline
 .7  & .084& .087& .093& .101& .113& .133& .167& .234& .389& .783&1.275\\
   & .206& .206& .206& .207& .208& .209& .212& .219& .233& .269& .314\\ \hline
.8  & .054& .058& .064& .072& .085& .106& .144& .218& .393& .874&1.551\\
   & .207& .207& .208& .208& .209& .210& .212& .216& .225& .252& .290\\ \hline
\end{tabular}

\begin{thebibliography}{70}
\bibitem{schild2} M. Bilenky, J.-L. Kneur, F.M. Renard, \& D. Schildknecht;\\
Nuclear Physics B 419 (1994) 240.
\bibitem{snowmass} T. Barklow {\em et.al}; hep-ph 9611454.
\bibitem{other} F. Boudjema; hep-ph 9701409. \\
G.J. Gounaris \& Costas G. Papadopoulos; hep-ph 9612378 \\
K. Hagiwara, T. Hatsukano, S. Ishiwara \& R. Szalapski;\\
Nuclear Physics B 496(1997) 66. 
\bibitem{pittau} Ezio Maina, Roberto Pittau, \& Marco Pizzio; \\
hep-ph 9709454 \& hep-ph 9710375.
\bibitem{schild1} M. Bilenky, J.-L. Kneur, F.M. Renard, \& D. Schildknecht; \\
Nuclear Physics B 409 (1993) 22.
\bibitem{Farhi} E. Farhi; Physical Review Letters 39 (1977) 1587.
\bibitem{NPB} K.J. Abraham \& Bodo Lampe; Nuclear Physics B 478 (1996) 507.
\bibitem{schmidt} Carl R. Schmidt; Physical Review D 54 (1996) 3250.
\bibitem{malaampi} M. Kuroda, F.M. Renard, \& D. Schildknecht; \\
Physics Letters B 183 (1987) 366 \\
C. Bilchak, M. Kuroda, \& D. Schildknecht; \\
Nuclear Physics B 299 (1987) 7. 
\bibitem{dawval} S. Dawson \& G. Valencia; Physical Review D 49 (1988) 2188 \\
A.A. Likhoded, T. Han, \& G. Valencia; \\
Physical Review D 53 (1996) 4811.
\end{thebibliography}
\end{document}